\documentclass[twocolumn,showpacs,superscriptaddress,amsmath,amssymb,prb]{revtex4}

\pdfoutput=1

\usepackage{amsmath,amsfonts,amssymb,bm}
\usepackage{dcolumn}
\usepackage[pdftex]{graphicx}
\usepackage{comment}
\usepackage{bm}

\begin{document}

\bibliographystyle{apsrev}	

\title{Phonon--assisted transitions from quantum dot excitons to cavity photons}

\author{Ulrich Hohenester}%
  \email{ulrich.hohenester@uni-graz.at}%
\affiliation{Institut f\"ur Physik,
  Karl--Franzens--Universit\"at Graz, Universit\"atsplatz 5,
  8010 Graz, Austria}

\author{Arne Laucht}
\author{Michael Kaniber}
\author{Norman~Hauke}
\author{Abbas Mohtashami}
\affiliation{Walter Schottky Institut, Technische Universit\"at M\"unchen,
  Am Coulombwall 3, 85748 Garching, Germany}

\author{Marek Seliger}
\affiliation{Institut f\"ur Physik,
  Karl--Franzens--Universit\"at Graz, Universit\"atsplatz 5,
  8010 Graz, Austria}

\author{Jonathan J. Finley}
\affiliation{Walter Schottky Institut, Technische Universit\"at M\"unchen,
  Am Coulombwall 3, 85748 Garching, Germany}

\pacs{73.21.La,42.50.Pq,63.22.-m,78.47.Cd}


\date{\today}

\begin{abstract}

For a single semiconductor quantum dot embedded in a microcavity, we theoretically and experimentally investigate phonon-assisted transitions between excitons and the cavity mode. Within the framework of the independent boson model we find that such transitions can be very efficient, even for relatively large exciton-cavity detunings of several millielectron volts.  Furthermore, we predict a strong detuning asymmetry for the exciton lifetime that vanishes for elevated lattice temperature. Our findings are corroborated by experiment, which turns out to be in good quantitative and qualitative agreement with theory. 

\end{abstract}

\maketitle


Quantum optical experiments using solid state quantum emitters have made a number of significant advances over the past decade. In particular, a series of manifestly `atom-like' properties have been identified for semiconductor quantum dots (QDs) such as a discrete optical spectrum, driven Rabi oscillations\cite{Zrenner:02,Ramsay:08}, dressed states\cite{Kroner:08,Muller:08} and, even, the optical control of single charge\cite{Zrenner:02,Ramsay:08} and spin\cite{Ramsay:08,Gerardot:08} excitations. It can be expected that established methods from quantum optics will continued to be applied to QDs, with good prospects for applications in quantum information science\cite{mabuchi:02} and for the development of novel devices, such as ultra-low threshold nano-lasers\cite{Nomura:06}. Many of the studies have involved the use of cavity QED\cite{gerard:98,reithmaier:04,hennessy:07,press:07,Kaniber:08a,Kaniber:08b,laucht:09} phenomena to control the spontaneous emission properties of the QD emitter by placing it in an optical nanocavity. This has opened the way to use the Purcell effect\cite{purcell:46} in the optical regime to build efficient sources of single photons\cite{Kaniber:08a,Kaniber:08b} and, even, observe single photon non-linearities in the strong coupling regime of the light matter interaction\cite{reithmaier:04,hennessy:07,press:07,Englund:07,laucht:09}. These phenomena are only expected to be active when the emitter ($\omega_{QD}$) and cavity ($\omega_{cav}$) frequencies are tuned close to resonance. In particular, the spontaneous emission lifetime is expected to reduce as the dot-cavity detuning ($\Delta\omega=\omega_{QD}-\omega_{cav}$) reduces, with a spectral dependence that reflects the cavity linewidth ($\kappa=Q/\omega_{cav}$).     
However, a number of recent studies have highlighted that QDs can non-resonantly couple to the cavity mode, even for detunings $\gg\hbar\kappa$.~\cite{hennessy:07,press:07,Kaniber:08b,winger:09} Such observations deviate strongly from the `artificial  atom' model and reflect the fact that the QD is far from an ideal two level system and, furthermore, it can couple to the degrees of freedom in the solid state environment. Such non-resonant dot-cavity coupling was shown to be crucial for a quantitative understanding of several experimental results.~\cite{laussy:08,laucht:09b,suffcynski:09} However, the microscopic mechanism continues to be somewhat controversially discussed in the community.\cite{Kaniber:08b,winger:09,suffcynski:09}\\
In this paper, we theoretically and experimentally investigate the role of phonon-assisted scattering between excitons and cavity modes within the framework of the independent boson Hamiltonian.~\cite{mahan:81,wilson-rae:02} We obtain an analytic expression for a scattering rate whose strength strongly depends on lattice temperature and detuning. Our results are corroborated by experimental results for a single quantum dot embedded in a photonic crystal nanocavity.


\emph{Theory}.---We start by considering the situation where a single exciton, confined in a semiconductor quantum dot, interacts with a single mode of a microcavity. Under appropriate conditions, we can describe the system as a genuine two-level system, where $|0\rangle$ denotes the cavity mode and $|1\rangle$ the exciton state. To account for the coupling of the exciton to the phonon environment, we introduce an effective Hamiltonian (we set $\hbar=1$)
\begin{equation}\label{eq:ham}
  \hat H_{\rm eff}=
  \hat H_0+\hat H'+\hat H_{\rm ph}+\hat H_{\rm ep}-\frac{i\hat\Gamma}2\,.
\end{equation}
Here, $\hat H_0=\Delta |1\rangle\langle 1|$ describes the uncoupled two-level system. $\Delta$ is the energy detuning between exciton and cavity, and we have set the cavity mode energy to zero. Throughout this paper we employ the rotating-wave approximation. $\hat H'=g\left(|1\rangle\langle 0|+|0\rangle\langle 1|\right)$ describes the electromagnetic coupling between the exciton and the cavity, which depends on the exciton dipole moment and the field distribution of the cavity mode.~\cite{andreani:99} The phonons and the exciton-phonon coupling are described by the \emph{independent boson Hamiltonian}~\cite{mahan:81,krummheuer:02,hohenester.jpb:07}
\begin{equation}\label{eq:hamib}
  \hat H_{\rm ph}+\hat H_{\rm ep}=
  \sum_\lambda \omega_\lambda 
  \hat a_\lambda^\dagger \hat a_\lambda^{\phantom\dagger}+
  \sum_\lambda g_\lambda\left(\hat a_\lambda^{\phantom\dagger}+
  \hat a_\lambda^\dagger\right) |1\rangle\langle 1|\,,
\end{equation}
with $\lambda$ denoting the different phonon modes with energy $\omega_\lambda$. $\hat a_\lambda^{\phantom\dagger}$ and $\hat a_\lambda^\dagger$ are the phonon field operators which obey the usual equal-time commutation relations for bosons. $g_\lambda$ are the exciton-phonon matrix elements, which depend on the material parameters of the host semiconductor and the exciton wavefunction.~\cite{krummheuer:02,hohenester.jpb:07} Eq.~\eqref{eq:hamib} describes a coupling where the exciton deforms the surrounding lattice and a polaron is formed, but no transitions to other quantum dot or cavity states are induced. Finally, $\hat\Gamma=\kappa|0\rangle\langle 0|+\gamma|1\rangle\langle 1|$ accounts for the radiative decay of the cavity and the quantum dot states. 

\begin{figure}
  \includegraphics[width=.7\columnwidth]{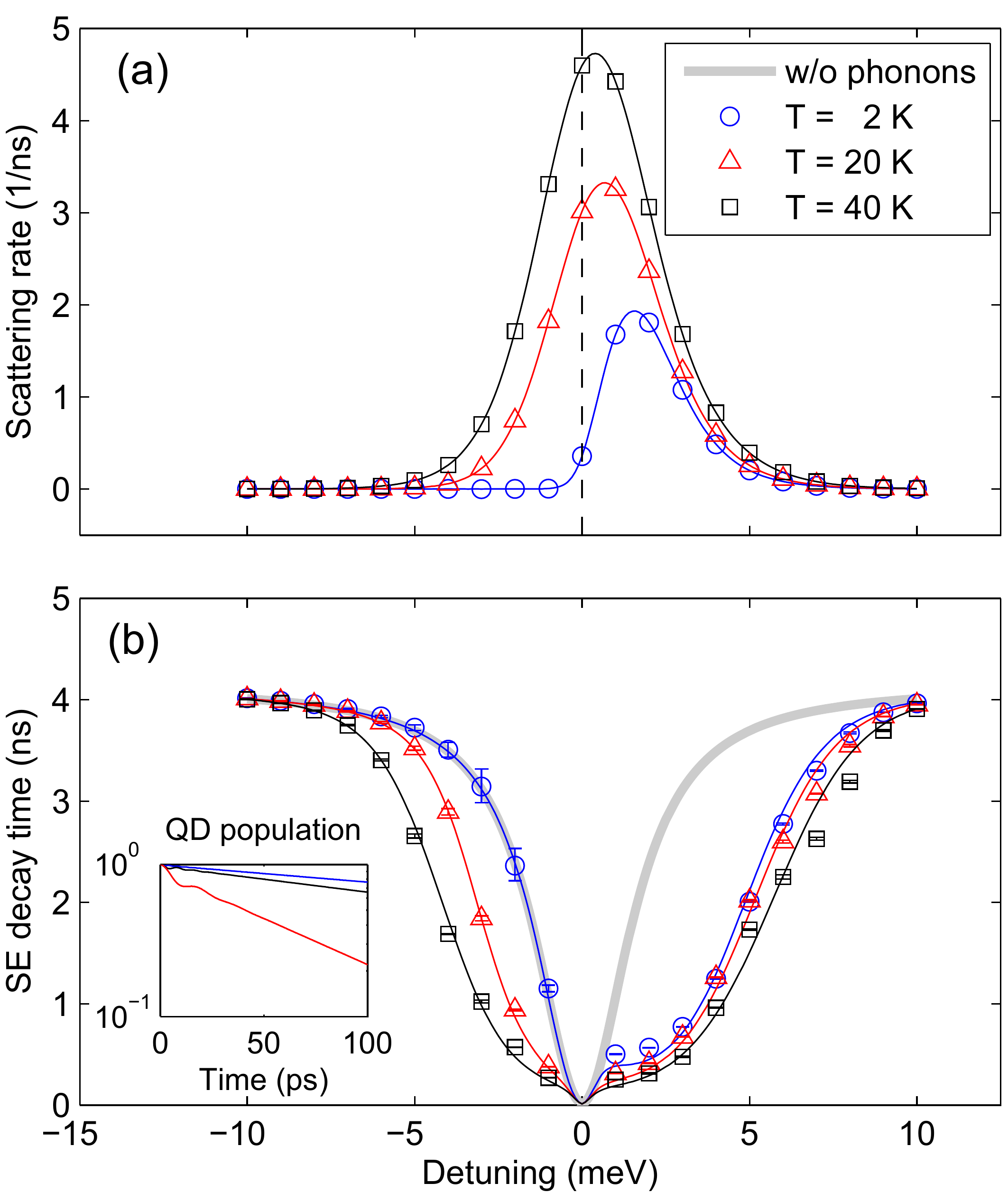}
\caption{(color online) (a) Phonon-assisted scattering rate from exciton to cavity as a function of exciton-cavity detuning, and for different temperatures. We use material parameters for GaAs and consider deformation potential and piezo electric interactions with acoustic phonons. For the electron and hole wavefunctions we assume Gaussians, with full-width of half maxima of 6 and 3.5 nm along the transversal and growth directions, respectively. For details of our model see Ref.~\onlinecite{hohenester.jpb:07}. (b) Full lines - spontaneous emission lifetime for the QD exciton as a function of temperature. We use $Q=10000$, $\hbar/\gamma=5$ ns, and $g=60\,\mu$eV. The symbols show results of our quantum-kinetic simulations, in good agreement with the rate equation model for the range of parameters investigated. The inset shows the calculated decay transients for detunings of $0.2$, $0.5$, and $1$ meV, in order of decreasing decay constant, and for $T=2$ K. }
\label{figtheory}
\end{figure}

Through $\hat H'$ the quantum dot state and the cavity mode become coupled, and new polariton modes are formed, which have partial exciton and partial cavity character.~\cite{walls:95,reithmaier:04,hennessy:07}. As we will show next, the phonon coupling opens a new decay channel, which is efficient also for relatively large detunings. Our starting point is the master equation in Born approximation~\cite{walls:95,wilson-rae:02}
\begin{equation}\label{eq:born}
  \frac{\partial\hat\rho(t)}{\partial t}
  \approx-\int_0^t\,\mbox{tr}_{\rm ph}\bigl\lgroup
  [\hat H'(t),[\hat H'(\tau),\hat\rho(t)\otimes \hat\rho_{\rm ph}]\,]
  \bigr\rgroup\,d\tau\,,
\end{equation}
where we use an interaction representation according to $\hat H_{\rm ib}=\hat H_0+\hat H_{\rm ph}+\hat H_{\rm ep}$. In Eq.~\eqref{eq:born}, we treat the exciton-cavity coupling $\hat H'$ as a ``perturbation'' and describe the exciton-phonon coupling without further approximation. Suppose that the system is initially in the exciton state $\hat\rho=|1\rangle\langle 1|$ and the phonons in thermal equilibrium. From Eq.~\eqref{eq:born}, we then obtain for the increase of population in the cavity 
\begin{equation}\label{eq:dotrho}
  \dot\rho_{00}\approx g^2\rho_{11}\int_0^\infty e^{i\Delta(\tau-t)}
  \mbox{tr}_{\rm ph}\left<1\right|\hat\rho_{\rm ph}e^{-i\hat H_{\rm ib}\tau}
  \left|1\right>\,d\tau+\mbox{c.c.}\,.
\end{equation}
This expression is of the form $\dot\rho_{00}=\Gamma_{10}\rho_{11}$ and, thus, can be associated with a scattering process. The memory kernel accounts for the buildup of a polaron state, and is governed by the correlation function 
\begin{equation}\label{eq:correlation}
  C(\tau)=\exp\bigl\lgroup-\sum_\lambda\left(\frac{g_\lambda}{\omega_\lambda}\right)^2
  \left[(\bar n_\lambda+1)e^{-i\omega_\lambda\tau}+\bar n_\lambda 
  e^{i\omega_\lambda\tau}\right]\bigr\rgroup
\end{equation}
This correlation function also determines pure phonon dephasing and the spectral lineshape in semiconductor quantum dots.~\cite{borri:01,krummheuer:02} $\hat n_\lambda$ is the occupation number of the phonons in thermal equilibrium. Then,
\begin{equation}\label{eq:scattrate}
  \Gamma_{10}=2g^2\,\Re e\int_0^\infty e^{i\Delta\tau}C(\tau)\,d\tau\,
\end{equation}
gives the scattering rate for a phonon-mediated transition from the exciton to the cavity mode. In a similar fashion, we can also determine the scattering rate $\Gamma_{01}$ for the reverse process, although this back-scattering is usually inefficient due to the strong losses $\kappa$ of the cavity.



\emph{Results}.---Figure 1(a) shows the calculated phonon-assisted scattering rate of Eq.~\eqref{eq:scattrate} as a function of detuning, and for different temperatures. One observes that at low temperatures the scattering is only efficient for positive (blue) detunings, i.e., when $\omega_{QD}\geq\omega_{cav}$. The corresponding scattering is associated with phonon emission. With increasing temperature phonon absorption becomes more important, and the scattering rate becomes more symmetric with respect to detunings. To estimate the importance of such phonon scatterings under realistic conditions, we consider the situation where initially the quantum dot is populated and subsequently decays through cavity and phonon couplings. We fully account for the electromagnetic coupling $\hat H'$ between exciton and cavity, and describe the radiative quantum dot and cavity losses through $\hat\Gamma$. The phonon scatterings are modeled in a master-equation approach of Lindblad form.~\cite{walls:95,laussy:08} We use typical values for state of the art cavity QED experiments, with the quality factor being $Q=10000$ ($\kappa=\hbar\omega_{QD}/Q$, where $\hbar\omega_{QD}\approx 1.3$ eV is the exciton transition energy), the exciton decay time $\hbar/\gamma=5$ ns, and the exciton-cavity coupling $g=60\,\mu$eV.~\cite{hennessy:07,Kaniber:08a} In the inset of Fig.~1(b) we present the decay transients calculated for different detunings. For the smallest detuning we observe oscillations of the quantum dot population at early times, which are a clear signature that the system operates in the strong coupling regime . In all cases we observe a mono-exponential decay at later times, which is associated with the radiative decay and phonon scatterings. Panel (b) of Fig.~1 shows the extracted decay times for different detunings and temperatures. At low temperatures we find that the decay is strongly enhanced at positive detunings, which is due to phonon emissions. This is seen most clearly by comparison with the thick gray line, that shows results of a simulation where phonon scatterings have been artificially neglected. Upon increasing the lattice temperature from $\sim2-40$~K phonon absorption gains importance, due to the thermal occupation of the relevant phonon modes, and the strong detuning asymmetry in the scattering rate is significantly weakened. Similar results are also obtained for smaller $Q$ values, where the dot-cavity system operates firmly in the weak-coupling regime of the light-matter interaction.  Thus, the results of the calculations presented in Fig.~1 are quite general and do not depend on the system operating in the strong or weak coupling regimes.

We additionally performed quantum-kinetic simulations for the coupled cavity-dot system in presence of phonon dephasing. Our approach is based on a density matrix description, including all one- and two-phonon assisted density matrices, and a surrogate Hamiltonian with a finite number of representative phonon modes.~\cite{foerstner:03a,hohenester.prl:04,hohenester.jpb:07} The symbols in Fig.~1(b) report results of these simulations. Throughout we observe an almost perfect agreement between the full simulations and the much simpler rate-equation approach.~\cite{remark.revival} This demonstrates that Eq.~\eqref{eq:scattrate} incorporates the pertinent physical mechanisms of phonon-assisted exciton-cavity scatterings.  Thus, we conclude that phonon mediated scattering into the cavity mode should: (i) enhance the bandwidth over which the spontaneous emission lifetime is influenced by the proximal cavity mode, and (ii) give rise to an asymmetry in the $\Delta\omega$ dependence of the spontaneous emission lifetime that should be observable for lattice temperatures $\leq40$~K.


\begin{figure}
  \centerline{\includegraphics[width=0.7\columnwidth]{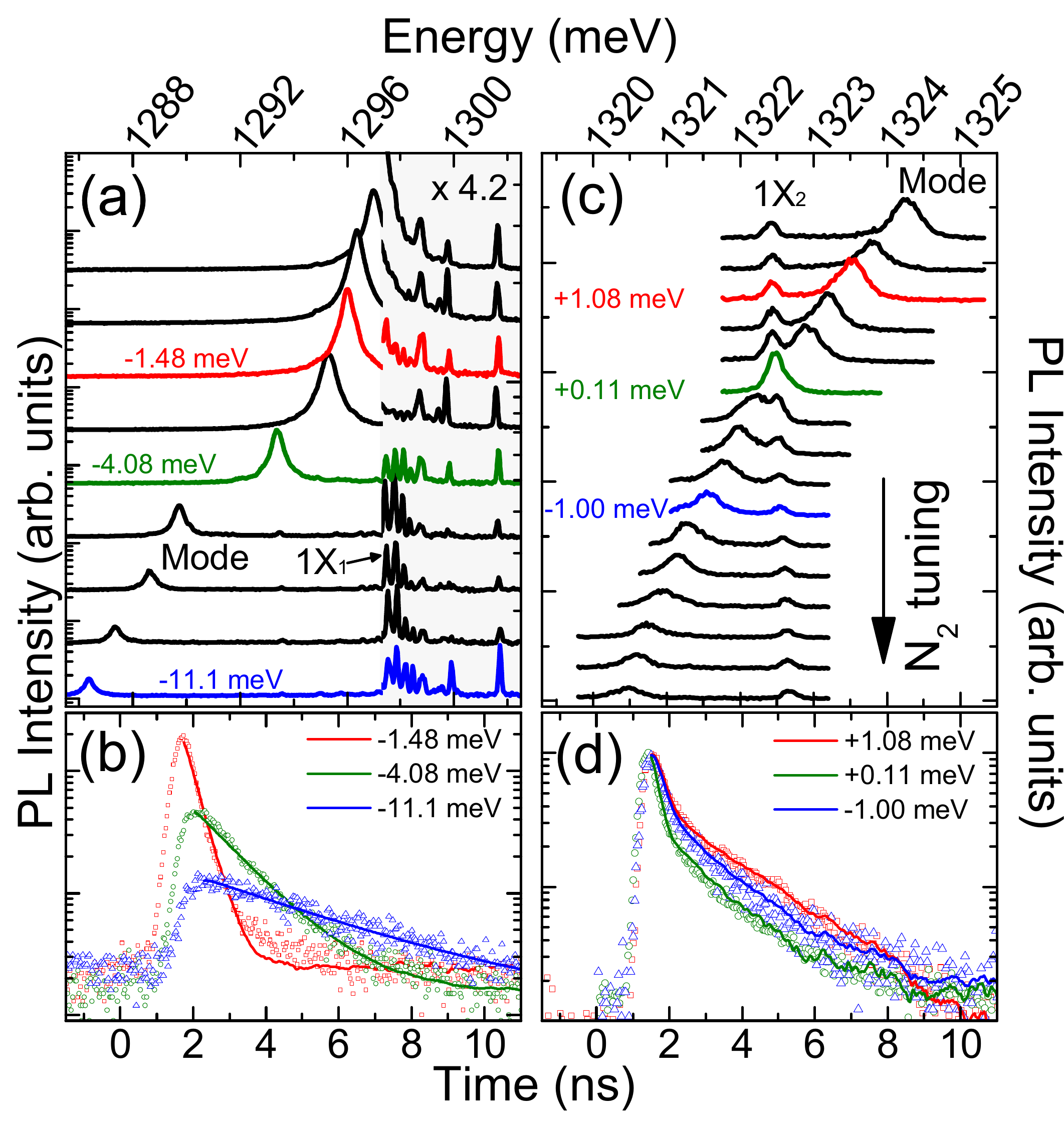}}
\caption{(color online) (a) Photoluminescence spectra of QD-cavity system I, with $Q=2900$ and $T=15$ K. The mode was spectrally tuned away from the QD s-shell to lower energies by adsorption of molecular nitrogen. (b) Decay transients of the indicated QD state for three different dot-cavity detunings. (c) Photoluminescence spectra of QD-cavity system II, with $Q=3200$ and $T=18$ K. The mode was spectrally tuned through a QD transition to lower energies by adsorption of molecular nitrogen. (d) Decay transients of the QD transition for three different quantum dot-cavity detunings. 
}
\label{figweak}
\end{figure}

\emph{Experiment}.---%
We corroborate our theoretical calculations by comparing with experiments. Here, we investigated the spontaneous emission dynamics of two different single QD--cavity systems, as a function of the exciton-cavity mode detuning. The samples consist of a single layer of InGaAs self-assembled quantum dots embedded at the center of a 180 nm thick GaAs membrane. Patterning via electron beam lithography and reactive ion etching results in the formation of a photonic crystal slab with an L3 defect nanocavity. Optical measurements were performed using confocal microscopy at low temperatures in a helium flow cryostat. The signal was spectrally analyzed by a $0.55$~m imaging monochromator and detected with a Si-based, liquid nitrogen cooled CCD detector. For time-resolved measurements, we used a fast silicon avalanche photodiode that provides a temporal resolution of $\sim100$~ps after deconvolution. We employed the technique of nitrogen deposition to systematically shift the cavity mode and vary $\Delta\omega$.~\cite{mosor:05} Furthemore, we perfomed control measurements to check that this tuning method has no direct influence on the intrinsic spontaneous emission dynamics of the QD itself.\\

Fig.~\ref{figweak}(a) shows a series of photoluminescence spectra recorded for QD-cavity system I as a function of successive nitrogen deposition steps.\footnote{The PL spectra were recorded with relatively strong excitation levels ($\sim10~W/cm^2$) in order to clearly observe the cavity mode, whilst the time resolved measurements presented later were performed with much weaker excitation power ($\leq1~W/cm^2$) in order to avoid state filling effects.} A number of single exciton-like transitions were observed in the vicinity of the QD s-shell interband transition and were identified by their linear power dependence.  We focused on one such prominent single exciton transition, labelled $1X_1$ on Fig.~2(a). For each nitrogen detuning step we observed a clear shift of the mode emission towards lower energies and away from the emission of $1X_1$ up to a maximum detuning of $\hbar\Delta\omega\sim 11$~meV. In Fig.~\ref{figweak}(b) we plot three representative decay transients recorded whilst detecting on $1X_1$. The solid lines on Fig.~2(b) are fits to the time resolved data convoluted with the instrument response function of the detection system.  All transients show a clear monoexponential decay with a decay time that becomes systematically slower as the cavity mode is detuned, precisely as expected due to the Purcell effect. However, as we will show below the cavity mode exerts an influence on the decay lifetime for detunings far beyond what would be expected due to the Purcell effect. 

Fig.~\ref{figweak}(c) shows a series of representative photoluminescence spectra recorded from QD-cavity system II. In this case, we conducted measurements for positive and negative detunings where the QD is at lower and higher energies than the mode ($-2$~meV $\leq\Delta\omega\leq2$~meV), respectively. However, the absolute range of detunings was lower as compared with system I. We excited the system in resonance with a higher energy cavity mode at 1433 meV and an excitation power density far below the saturation of the exciton ($\sim1$ W/cm${}^2$).~\cite{kaniber:09} Three representative decay transients are plotted in Fig.~\ref{figweak}(d). In contrast to system I, we observe a clear biexponential decay for all detunings, with both fast ($\tau_f$) and slow ($\tau_s$) decay components. These two lifetimes can be attributed to the two fine structure components of the neutral exciton that couple with a different strength to the linearly polarized cavity mode. \footnote{This is inevitable for these systems since the L3 cavity axis was not intentionally aligned with the [110] crystal axis and the neutral exciton consists of a linearly polarized doublet with a fine structure splitting below our spectral resolution of $60$ $\mu$eV.} A comparison of the three decay transients presented in Fig.~2(d) reveals that both $\tau_f$ and slow $\tau_s$ are minimum close to zero detuning, as expected due to the Purcell effect.~\cite{purcell:46} The other two decay transients selected for presentation in Fig.~\ref{figweak}(b) were recorded at $\hbar\Delta\omega=+1$~meV and $-1$~meV, respectively and exibit markedly differing lifetimes despite the similar magnitude of the detuning. This observation already hints at an asymmetry in the lifetimes for positive and negative detunings.\\

\begin{figure}
  \centerline{\includegraphics[width=0.7\columnwidth]{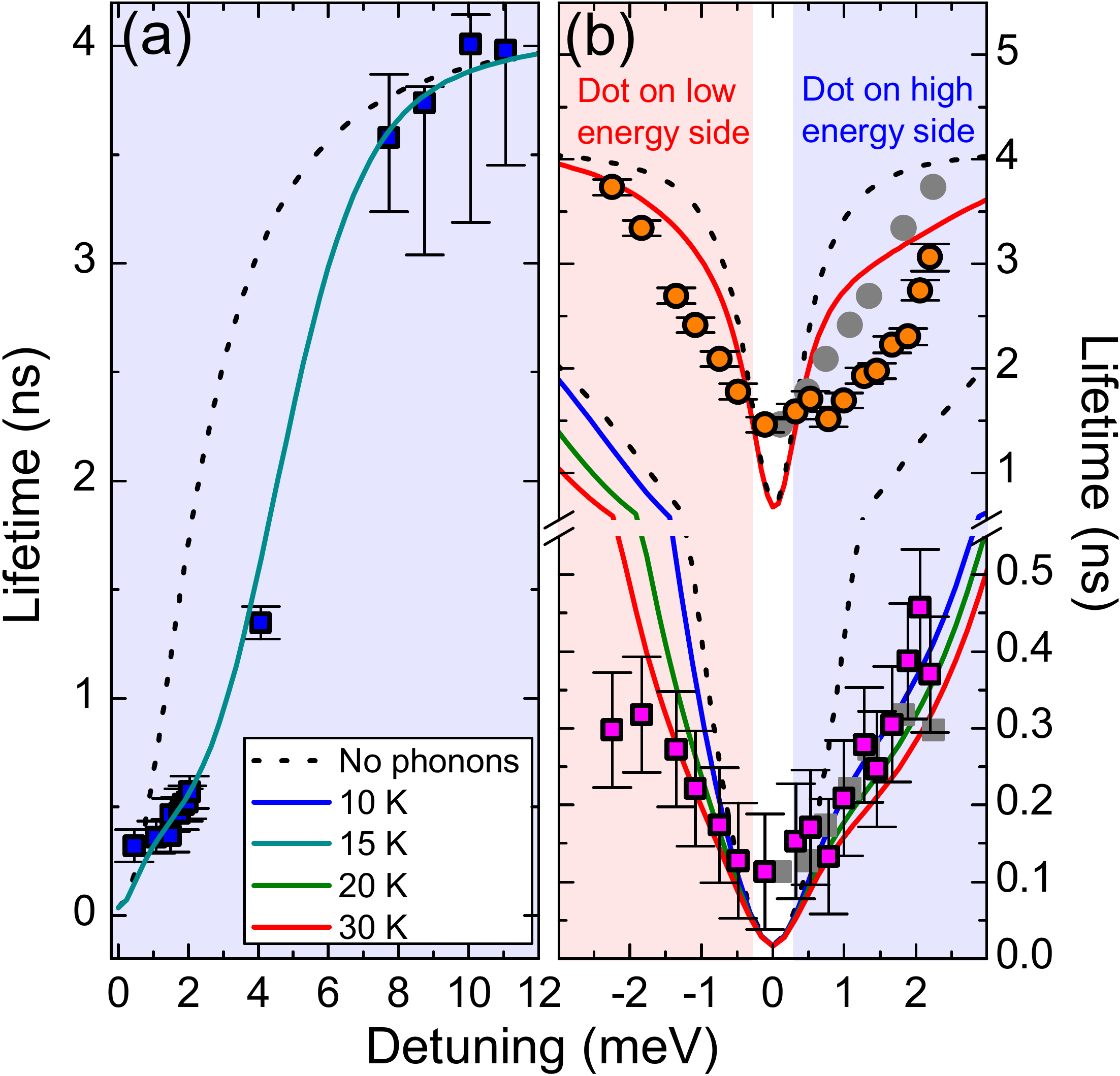}}
\caption{(color online) (a) Extracted decay times of QD-cavity system I as a function of detuning. The squares correspond to the $1X_1$ state. The black dotted line and the cyan solid line are theoretical decay times obtained from calculations with the phonon coupling turned off and a phonon bath at 15 K temperature, respectively. (b) Extracted decay times of QD-cavity system II as a function of detuning. The squares and circles correspond to the fast and the slow decay component, respectively. The gray traces are shown for better comparison, and display the data for negative detunings mirrored on the positive axis. The lines are calculations of the system without phonon coupling and for different temperatures (10 K, 20 K, 30 K). Note that the measured lifetimes in resonance are limited by the temporal resolution of the detection system of $\sim150$~ps after deconvolution.
}
\label{figfit}
\end{figure}

In Fig.~\ref{figfit}(a) we summarize all measured lifetimes for QD-cavity system-I as a function of $\Delta\omega$. We performed simulations using the experimentally measured quality factor $Q=2900$ and $g=45$ $\mu$eV. When the phonon assisted dot-cavity coupling is artificially neglected (black dotted line) the experimental data (squares) cannot be reproduced and, furthermore, the dip in $\tau(\Delta\omega)$ is much narrower than is observed experimentally. However, when coupling to a phonon bath with a temeprature of 15~K is switched on (solid cyan line), the calculation accounts very well for the measured data.  Similarly, in Fig.~\ref{figfit}(b) we summarize the information extracted from the decay transients of QD-cavity system II. We plot the measured lifetimes $\tau_s$ and $\tau_f$ with circles and squares, respectively. The gray traces are the data points for negative detunings mirrored onto the positive axis to facilitate a direct comparison. For the slow decay component, we clearly observe a pronounced asymmetry in the data as a function of detuning, with the measured lifetimes being generally shorter for positive detunings as compared to negative detunings. The fast decay component is more symmetric with respect to detunings but also reveals a dip with a width that can only be accounted for when phonon assisted scattering into the cavity is included.  To see this, we calculated the lifetimes without coupling to phonons (black dotted line), and coupling to a phonon bath of 10~K (blue solid line), 20~K (green solid line) and 30~K (red solid line). We used $Q=3200$ and $g=60$ $\mu$eV as obtained from the optical measurements. It is obvious that coupling to the phonon bath is needed to correctly represent the dip in the measured lifetime, which underlines the role of phonons as feeding channel of the cavity mode at small detunings. As for the slow component, we consider a level scheme with two fine structure components of the exciton, where $g=10$ $\mu$eV for the weakly coupled component. In Fig.~\ref{figfit}(b) we present results of simulations with coupling to a phonon bath at $30$~K (red solid line) and without phonon scatterings (black dotted line). Whilst the overall agreement with experiment is not fully satisfactory, our calculations qualitatively reproduce the asymmetric detuning dependence in good qualitative agreement with experiment. 


In conclusion, we have investigated phonon-assisted transitions between excitons in quantum dots and photons in a nanocavity. These transitions are efficient over a wide range of detunings, and strongly depend on temperature. Our theoretical predictions are qualitatively and quantiatively corroborated by experiment.

We acknowledge financial support of the DFG via the SFB 631, Teilprojekt B3 and the German Excellence Initiative via NIM.


\begin{thebibliography}{32}
\bibitem[{rem()}]{remark.revival}
\bibinfo{note}{The small discrepancies between the quantum-kinetic and
  rate-equation simulations are attributed to the too small number of about
  hundred phonon modes in our simulation, which induce artificial phonon
  revivals in the simulated decay transients.}
\end{thebibliography}
\end{document}